\documentclass[aps,prd,preprint,groupedaddress,showpacs,amsfonts,amssymb,amsmath,floatfix]{revtex4}
\usepackage{bm}
\usepackage{graphicx}

\begin{document}

\title{Differing Calculations of the Response of Matter-wave
Interferometers to Gravitational Waves}

\author{A.~D.~Speliotopoulos}

\email{adspelio@uclink.berkeley.edu}

\affiliation{Department of Physics, University of California,
Berkeley, CA 94720-7300}

\author{Raymond Y. Chiao}

\email{chiao@physics.berkeley.edu}

\homepage{physics.berkeley.edu/research/chiao/}
{}
\affiliation{Department of Physics, University of California,
Berkeley, CA 94720-7300}

\date{June 23, 2004}

\begin{abstract}

There now exists in the literature two different expressions
for the phase shift of a matter-wave interferometer caused by the
passage of a gravitation wave. The first, a commonly accepted
expression that was first derived in the 1970s, is based on the
traditional geodesic equation of motion (EOM) for a test particle. The
second, a more recently derived expression, is based on the geodesic
deviation EOM. The power-law dependence on the frequency of the
gravitational wave for both expressions for the phase shift is
different, which indicates fundamental differences in the physics on
which these calculations are based. Here we compare the two
approaches by presenting a series of side-by-side calculations of the
phase shift for one specific matter-wave-interferometer
configuration that uses atoms as the interfering particle. By looking
at the low-frequency limit of the different expressions for the phase
shift obtained, we find that the phase shift calculated via the
geodesic deviation EOM is correct, and the ones calculated via the
geodesic EOM are not.

\end{abstract}

\pacs{39.20.+q, 04.20.Cv, 04.30.Nk}

\maketitle

% If in two-column mode, this environment will change to single-column
% format so that long equations can be displayed. Use
%\begin{widetext}
%\end{widetext}

\section{Introduction}

In two recent papers \cite{nmMIGO, CrysMIGO}, a new calculation of
the phase shift of a matter-wave interferometer caused by the
passage of a gravitational wave was presented. This calculation
was based not  on the usual geodesic equation of motion (EOM), but
rather on the geodesic \textit{deviation} EOM. Contrary to
expectations, these phase shifts have a form that is very much
different from those calculated previously in the literature
\cite{Linet-Tourrenc, Stodolsky, Papini, Borde1,Borde2, Borde3,
Alsing}. While a direct comparison of the various expressions for
the phase shift is difficult because of the different
configurations of the matter-wave interferometers used for their
calculations, it was noted in \cite{nmMIGO} that the power-law
dependence on the frequency of the gravitational wave, in both the
low- and high-frequency limit, of the expressions calculated in
\cite{nmMIGO, CrysMIGO} are at times diametrically opposite to
those calculated in \cite{Linet-Tourrenc, Stodolsky, Papini,
Borde1,Borde2, Borde3, Alsing}. Consequently, we expect that the
differences between these expressions are \textit{not} due simply
to differences in the configurations of the interferometers
chosen, but are rather caused by fundamental differences in the
physics on which their derivations are based.

In this paper our objective is to demonstrate that these
fundamental differences do exist by presenting a series of
calculations that follows the two basic approaches in the literature used
to determine the phase shift caused by a gravitational wave. This will
be done for the \textit{same} configuration of matter-wave
interferometer, and will allow us to compare the different
expressions we obtain for the phase shift on an equal footing. We then
determine their validity by comparing each of them, in the appropriate
limit, with the well-known expression for the phase shift caused by
\textit{stationary} sources of spacetime curvature \cite{Anan, GRG,
Papini}. The focus in this paper will be on a \textit{theoretical}
study of the underlying physics; we leave issues of the practicality
and feasibility of constructing the interferometer to \cite{nmMIGO,
CrysMIGO}.

The various approaches taken to determine the phase shift of a matter-wave
interferometer caused by a gravitational wave can be divided into two
distinctly separate classes. The class of approaches that is more
often found in the literature \cite{Linet-Tourrenc, Stodolsky, Papini,
Borde1,Borde2, Borde3, Alsing} is done in transverse-traceless (TT)
coordinates (see Sec.~9.2.3 of \cite{Thorne}), either explicitly or
implicitly. The TT coordinates are one specific choice of reference frame
where the gravitation wave is in the TT gauge. In this frame the number of
components of the deviations of the flat spacetime metric
$h_{\mu\nu}$ caused by the gravitational wave equals the number of
physical degrees of freedom of the gravitational wave. Consequently,
$\partial^\mu h_{\mu\nu}$, $h_{0\mu} = 0$, and $h_\mu^\mu=0$. In
addition, the mirrors and the beam splitters are
at fixed coordinate values in these coordinates, and they do not
move. These coordinates have been used to study the properties of
light-wave-interferometry-based detectors of
gravitational waves as well \cite{Hellings}. To calculate the phase
shift for matter-wave interferometers, it is
combined with either the geodesic EOM (and its action), if the
calculation is done at the quantum-mechanical level
\cite{Linet-Tourrenc, Alsing}, or the Lagrangian (and the WKB or
stationary phase approximation) for a quantum field in curved
spacetime, if the calculation is done at the quantum field-theoretic
level \cite{Papini, Borde1, Borde2, Borde3}. Not surprisingly,
irrespective of the level of sophistication of the approach used, the
general expressions calculated for the phase shift have the same final
form.

The other class of approaches to calculating the phase shift is taken by
\cite{nmMIGO, CrysMIGO}, and is done in Thorne's ``proper
reference frame''. They follow the classical analysis in
\cite{Thorne, Thorne1983, MTW} for light-wave-based
interferometers. In this frame, the positions and velocities of
test particles are measured relative to the worldline of an
observer, and follow the motion of the observer in spacetime, in
much the same way that the Fermi-normal \cite{Fermi}, Fermi-Walker
coordinates \cite{Synge}, and the recently constructed general
laboratory frame \cite{GLF} do. Consequently, the motion of the
test particle in this frame is \textit{not} described by the usual
geodesic equation of motion (EOM), but rather by the geodesic
\textit{deviation} EOM (see Eq.~(2) of \cite{Thorne}), which was
first applied to the analysis of physical systems in
\cite{Synge1926, LC}. As in \cite{Thorne}, the TT gauge was taken for
the gravitational wave. The action (see \cite{Synge1935, ADS1995,
GLF}) for the geodesic deviation EOM was then used in conjunction with
the quasi-classical approximation of the Feynman-path-integral
formulation of quantum mechanics. This approximation is the
Lagrangian form of the WKB approximation in the Schr\"odinger
representation of quantum mechanics, and in \cite{nmMIGO, CrysMIGO} it
was applied to the calculation of the phase shift of the two
matter-wave-interferometer configurations.

The phase shift of a matter-wave interferometer is a physically
measurable quantity. From the principle of general
covariance, it should not matter whether the TT coordinates, or
whether the proper reference frame is used in its calculation
\cite{Thorne, Unruh-Weiss}. Indeed, Thorne states the following:
\begin{quote}
``If $L<<\lambdabar$ then the [gravitational wave] detector can be
contained entirely in the proper reference frame of its center, and
the analysis can be performed using non-relativistic concepts
augmented by the quadrupolar gravity-wave force field (3), (5). If one
prefers, of course, one instead can analyze the detector in TT
coordinates using general relativistic concepts and the spacetime
metric (8). The two analyses are guaranteed to give the same
predictions for the detector's performance, \textit{unless errors are
made} [emphasis added].''
(From page 400 of \cite{Thorne}.)
\end{quote}
Evidently, errors have indeed been made, as we shall see, since the
resulting calculations often contradict each other for the same
interferometer.  The power-law dependence on frequency of the phase
shifts calculated in \cite{Linet-Tourrenc,Stodolsky, Papini,
Borde1,Borde2, Borde3, Alsing} using TT coordinates are
fundamentally different from that of the phase shifts calculated in
\cite{nmMIGO, CrysMIGO} using the proper reference frame. Since the
principle of general covariance must hold, this means that either (a) the
calculations in \cite{nmMIGO, CrysMIGO} are wrong, (b) the
calculations in \cite{Linet-Tourrenc,Stodolsky, Papini,
Borde1,Borde2, Borde3, Alsing} are wrong, or (c) all of the
calculations of the phase shifts are wrong.

To determine which one of these three possibilities is correct, we
will present side by side calculations of the phase shift caused by a
gravitational wave for one specific matter-wave-interferometer
configuration using the following three approaches. The first
calculation uses the quasi-classical approximation in the TT
coordinates and geodesic action, and is done at the nonrelativistic,
quantum-mechanics level. It follows, as much as possible, the line of
analysis given in \cite{Stodolsky}. The second calculation is also
done at the nonrelativistic, quantum-mechanics level using the same
approximation, but now in the proper reference frame, and uses the
geodesic \textit{deviation} action. It follows the line of analysis
given in \cite{nmMIGO}.

These two calculations would seem to exhaust the two different classes
of approaches currently found in the literature. That a third
calculation of the phase shift is needed is because of an error
in \cite{Linet-Tourrenc, Stodolsky, Papini, Borde1,Borde2, Borde3,
Alsing}. Either because changes in the kinetic energy term in the
action for the atom in \cite{Stodolsky, Alsing} were neglected, or
because the turning points of the atom's phase at the interferometer's
mirrors in the WKB-type of approach in \cite{Linet-Tourrenc, Papini,
Borde1, Borde2, Borde3} were omitted, the end result is that important
boundary terms are not included in the final expression for the phase
shift in \cite{Linet-Tourrenc, Stodolsky, Papini, Borde1,Borde2, Borde3,
Alsing}. For the sake of completeness, and to compare the calculations in
\cite{nmMIGO, CrysMIGO} with these previous calculations in the
literature, this necessitates a \textit{third} calculation of the
phase shift, which will result in yet another expression for the phase
shift based on the geodesic EOM.

We find that all three expressions for the phase shift calculated
here are different from one another. Although there are currently
no \textit{direct} experimental checks that can used to
determine which one of these three expressions for the phase
shift is correct, an \textit{indirect} experimental check can be
used by taking the low-frequency limit.

All three expressions for the phase shift depend on the transit
time of the atom through the interferometer, and the period
of the gravitational wave. If the transit time of the atom is much
shorter than the period, then, with respect to the atom, the
gravitational wave does not oscillate appreciably as it traverses
the interferometer. The gravitational wave will contribute a local
Riemannian curvature to the spacetime that is effectively static.
The phase shift of matter-interferometers caused by static (as
well as stationary) sources of curvature is well established
\cite{Anan, GRG, Papini}, and has been shown to be proportional to
components of the Riemann curvature tensor. Just as importantly, this
phase shift can be calculated using only Newtonian gravity [Eq.~(2.6) of
\cite{Anan}]. The phase shift due to the Newtonian gravitational
potential\textemdash through the acceleration due to Earth's
gravity $g$\textemdash has been measured, first by Collela,
Overhauser and Werner \cite{COW}, and more recently by Chu and
Kasevich \cite{ChuKas}. These more recent measurements of the phase
shift induced by $g$ are sensitive enough that effect of
\textit{spatial variations} in $g$\textemdash which are proportional to the
static Riemann curvature of the spacetime caused by the Earth\textemdash
on the phase shift can be discerned. Gradients in $g$ were seen
through \textit{changes} in the phase shift of an atomic fountain when its
vertical position was varied \cite{PCC}. More recently, they were seen
through a differential measurement of the phase shifts of two atomic
fountain matter-wave interferometers operating in tandem
\cite{Kas}. Therefore, while the direct measurement of the phase shift
of a matter-wave interferometer caused by a static Riemann curvature
tensor has not been made, its discovery is expected. Indeed,
indications are that the phase shift induced by the static Riemann
curvature of the Earth has already been seen \cite{PrivateKas}.

As a consequence of the above, in the limit of low-frequency
gravitational waves the phase shift calculated for the matter-wave
interferometer \textit{must} approach the well-known static result,
and be proportional to the curvature tensor. We will see that in this
limit, only the phase shift calculated along the lines of \cite{nmMIGO,
CrysMIGO} has the correct form. The other two do not, and \textit{must
be wrong}, since they do not yield the correct static limit.

Our focus here is on the phase shift that a \textit{gravitational
wave} will induce on a matter-wave interferometer. We will
thus neglect the effects of all other gravitational
effects\textemdash such as the acceleration due to Earth's gravity,
and the local curvature of spacetime due to \textit{stationary}
sources such as the Earth, Sun and Moon\textemdash in addition to the
phase shift due to the Sagnac effect caused by the Earth's
rotation. We have shown in \cite{GLF} that for these types of
stationary gravitational effects the description of the motion of test
particles in the general laboratory frame\textemdash of which the
proper reference frame is a special case\textemdash reduces to the usual
geodesic EOM description, and there is no controversy. The phase
shifts caused by these stationary effects calculated in
either the proper reference frame or the TT coordinates will be the
same.

We will in this paper use the same notation the is commonly found
in the literature \cite{nmMIGO, CrysMIGO, Linet-Tourrenc,
Stodolsky, Papini, Borde1,Borde2, Borde3, Alsing}. In particular,
it follows the notation in \cite{Thorne}. While this has the
benefit of consistency and simplicity, it means that we do not
separate with notation the coordinates used for the TT coordinates
from the coordinate used for the proper reference frame. The
coordinates used in the two reference frames have different
physical meanings, which we will make clear in the next section.
These differences are outlined in detail in \cite{GLF}.

\section{Phase-Shift Calculations}

In this section we will present two of the three calculations of
the phase shift mentioned in the introduction. The first
calculation will be done in the TT coordinates, and will be based
on the geodesic EOM. The second calculation will be done in the
proper reference frame, and will be based on the geodesic
deviation EOM. A third calculation, following the approach in the
literature, will be done in the next section where comparisons
between the different results of the calculations are made. All
three calculations are done for gravitational waves in the
long-wavelength limit for the matter-wave interferometer
configuration shown in Fig.~$\ref{H-V-MIGO}$; this configuration
is described in detail in \textbf{appendix A}. To calculate the
phase shifts, we will use the quasi-classical approximation of the
Feynman path integral representation of quantum mechanics. This
approximation is the Lagrangian version of the WKB approximation,
or stationary phase approximation, in the Schr\"odinger
representation of quantum mechanics, and will be reviewed in
\textbf{appendix A} as well. The presentation in this section is
done in detail to ensure that all the underlying assumptions,
approximations, and subtleties in our analysis are readily
apparent.

\subsection{Geodesic-EOM-based calculation of the phase shift}

\subsubsection{Classical Dynamics in TT Coordinates}

\begin{figure}[ptb]
\begin{center}
  \includegraphics[angle = 270, width=0.88\textwidth]{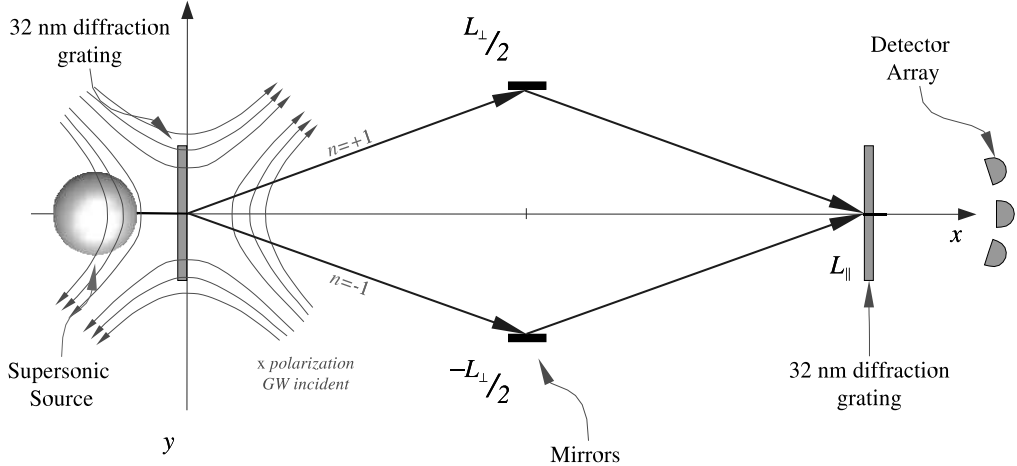}
\end{center}
\caption{Schematic of the matter-wave interferometer configuration
considered in this paper with a gravitational wave incident normal to
the plane of the interferometer shown. Only the $\times$ polarization
contributes to the phase shift; the $+$ polarization does
not. Diffraction orders other than $n=\pm1$ are not used, and are left
out for clarity. }
\label{H-V-MIGO}
\end{figure}

Consider a gravitational wave in the long-wavelength limit with
an amplitude $h_{\mu\nu}(t)$ incident
perpendicularly to the plane of the interferometer in
Fig.~$\ref{H-V-MIGO}$. As usual, Greek indices run from $0$ to $4$,
and Latin indices run from $1$ to $3$. The gravitational wave is
represented by the tensor $h_{\mu\nu}$ as a perturbation on the flat
spacetime metric $\eta_{\mu\nu}$, so that the total metric of the
spacetime is $g_{\mu\nu}=\eta_{\mu\nu}+h_{\mu\nu}$. We have chosen the
signature of $g_{\mu\nu}$ to be $(-1,1,1,1)$. In the TT-gauge, a
coordinate system is chosen so that in these coordinates,
$\partial^\mu h_{\mu\nu}=0$, $h_{0\mu}=0$, and $h_\mu^\mu=0$; the only
nonvanishing components of $h_{\mu\nu}$ are thus $h_{ij}$.

One choice of coordinates where the TT gauge is realized is the TT
coordinates (see \cite{Thorne} and the general analysis in
\cite{Wald}). In these coordinates the action for the atom with mass
$m$ used in Fig.~$\ref{H-V-MIGO}$ is the usual geodesic action in the
nonrelativistic limit,
\begin{equation}
S_{geo} =
-m\int\left(-\frac{dx^\mu}{dt}\frac{dx^\nu}{dt}g_{\mu\nu}\right)^{1/2}
dt \approx -mc^2t +\frac{1}{2}m \int\left[v^iv^j(\delta_{ij} +
  h_{ij})\right]dt,
\label{geoaction}
\end{equation}
where the subscript ``geo'' stands for ``geodesic''.
Since the rest mass $m c^2$ only shifts the Lagrangian for the atom by
a constant, we drop this term from $S_{geo}$. The nonrelativistic
Hamiltonian for the motion is then,
\begin{equation}
H_{geo} = \frac{\vec{p}^{\>2}}{2m} - \frac{1}{2m} p^ip^jh_{ij},
\label{geoham}
\end{equation}
which results in the usual geodesic equation of motion
\begin{equation}
\frac{d^2x^i}{dt^2} = -2\Gamma^i_{0j}\frac{dx^j}{dt} =
-\dot{h}^i_j\frac{dx^j}{dt}.
\label{geoeqn}
\end{equation}
Here $\Gamma^\alpha_{\mu\nu}$ is Christoffel symbol for the spacetime;
for gravitational waves in the TT gauge, $\Gamma^i_{0j}=\dot{h}^i_j/2$.

To calculate the phase shift of Fig.~$\ref{H-V-MIGO}$, we follow
the description in \textbf{appendix A}, and consider the phase
shift due to the $r$th atom emitted from the atomic source, which
at a time $t_r$ is diffracted by the initial beam splitter. There
are two possible classical paths for the atom through the
interferometer. In the absence of gravitational waves, these are
straight-line paths $\vec{x}_0(t)$ (see \textbf{appendix A}), which
will be shifted when a gravitational wave is present. These shifts are
expected to be very small, however, and we can use perturbation theory
to solve for them. We thus take $\vec{x}(t) = \vec{x}_0(t)
+ \vec{x}_1(t)$, and $\vec{v}(t) = \vec{v}_0(t) + \vec{v}_1(t)$,
where $\vec{x}_1(t)$ and $\vec{v}_1(t)$ are perturbations to the
atom's path $\vec{x}_0(t)$ and velocity $\vec{v}_0(t)$,
respectively; we will keep only terms linear in $\vec{v}_1(t)$. To
first order in $h_{ij}$, Eq.~$(\ref{geoeqn})$ becomes
\begin{equation}
\frac{dv_1^i}{dt} = -\dot{h}^i_j v_0^j.
\label{pert-geo}
\end{equation}
Once $\vec{x}_0(t)$ is determined, $v_1(t)$ is solved for by dividing
the trajectory of an atom into two parts: from the initial beam
splitter to the mirror, and then from the mirror to the final beam
splitter. The two paths are then joined with the appropriate boundary
conditions at the mirror.

We begin by determining $\vec{x}_0(t)$.

\subsubsection{Trajectories in the absence of gravitational waves}

To determine the unperturbed path of the atoms through the
interferometer\textemdash which is the same no matter which approach
is used to calculate the phase shifts\textemdash consider first the
path of the atom that is diffracted into the $n=+1$
diffraction order. We will assume that the velocity
of the atoms is high enough that we can ignore the effect of Earth's
gravity on the atom. It is then clear that for
$t_r<t<t_r+T/2$\textemdash during which it travels between the initial
beam splitter and the first mirror in a time $T/2$\textemdash the path
of the atom is
\begin{eqnarray}
v_{0x}=v_{\Vert},&\qquad& x_{0}(t)=v_{\Vert}(t-t_r),
\nonumber
\\
v_{0y}=v_{\bot},&\qquad& y_{0}(t)=v_{\bot}(t-t_r),
\label{firsthalf}
\end{eqnarray}
where $v_\|$ and $v_\bot$ are $x$ and $y$ components of the
velocity of the atom immediately after it leaves the initial beam
splitter. At $t_r+T/2$ the atom is reflected off the hard-wall
mirror, and from the general expression of the boundary condition
for the atom at the mirror Eq.~$(\ref{bc})$,
$v_{0x}(t_r+T^{+}/2)=v_{0x}(t_r+T^{-}/2)=v_\|$, while
$v_{0y}(t_r+T^{+}/2)=-v_{0y}(t_r+T^{-}/2)=-v_\bot$. (The
superscripts $\pm$ denote $\lim_{\epsilon\to 0} t_r + T/2\pm
\epsilon$ respectively.) For the subsequent time interval
$t_r+T/2<t<t_r+T$, the atom's path is
\begin{eqnarray}
v_{0x}=v_{\Vert}, &\qquad& x_{0}(t)=v_{\Vert}(t-t_r),
\nonumber
\\
v_{0y}=-v_\bot,& \qquad & y_0(t)=L_\bot-v_\bot(t-t_r),
\label{secondhalf}
\end{eqnarray}
during which it travels between the mirror and the final beam
splitter. It is clear that $L_{\bot}=v_{\bot}T$ and
$L_{\Vert}=v_{\Vert}T$, and $T$ is the total transit time of the atom
through the interferometer. The unperturbed path for the atom in the
$n=-1$ diffraction order is similar to Eqs.~$(\ref{firsthalf})$ and
$(\ref{secondhalf})$, and is obtained by taking $v_\bot\to -v_\bot$ and
$L_\bot\to-L_\bot$.

\subsubsection{Boundary conditions}

To determine the boundary conditions for $\vec{v}_1$ in
Eq.~$(\ref{pert-geo})$, we note that from Eq.~$(\ref{geoeqn})$ the
acceleration of a test particle for the geodesic EOM is proportional
to its velocity. As such, a test particle which was initially at rest
will stay at rest, \textit{even in the presence of a gravitational
wave}. This led to the following conclusion by Stodolsky:
\begin{quote}
``In this discussion [of matter-wave interferometry] we have assumed
that the parts of the apparatus [interferometer] stay at fixed
coordinate values. This can be achieved if these parts are free
masses, since a metric of the gravity wave type (with no ``0''
components) does not affect the positions of nonmoving bodies.'' (From
page 399 of \cite{Stodolsky}.)
\end{quote}
Thus, in the TT coordinates, the positions of beam splitters and the
mirrors do not change when a gravitational wave passes through the system.

The velocity of the atoms travelling between the supersonic source
and the initial beam splitter, on the other hand, \textit{will} be
affected by the gravitational wave. Since the interferometer in
Fig.~$\ref{H-V-MIGO}$ is \textit{balanced}, only
\textit{differences} in the atom's trajectory along the two paths
through it will matter. We can thus neglect any shifts in the atom's
velocity \textit{before} it reaches the beam splitter. And, since
the initial beam splitter is not moved by the gravitational wave, $v_{0x}(t_r)=v_\|$
while $v_{0y}(t_r)=v_\bot$; thus we take $\vec{v}_1(t_r)=0$.

For the boundary conditions at the mirror, we note that like the beam
splitter, the positions of the mirrors in the interferometer also do
not change when a gravitational wave is present. Thus the boundary
conditions for $\vec{v}_1(t)$ across the hard-wall mirrors is
\begin{equation}
v_{1x}(t_r+T^{+}/2) =v_{1x}(t_r+T^{-}/2), \qquad v_{1y}(t_r+T^{+}/2) =
- v_{1y}(t_r+T^{-}/2).
\label{pertGeobc}
\end{equation}

Integrating Eq.~$(\ref{pert-geo})$ is now straightforward, and for
$t_r<t<t_r+T/2$,
\begin{eqnarray}
v_{1x}(t) &=& -v_\| h_{xx}(t)- v_\bot h_{xy}(t)+v_\| h_{xx}(t_r)+
v_\bot h_{xy}(t_r),
\nonumber
\\
v_{1y}(t) &=& -v_\| h_{xy}(t)- v_\bot h_{yy}(t)+v_\| h_{xy}(t_r)+
v_\bot h_{yy}(t_r),
\label{geosolutions-first}
\end{eqnarray}
while for $t_r+T/2<t<t_r+T$,
\begin{eqnarray}
v_{1x}(t) &=& -v_\| h_{xx}(t) + v_\bot h_{xy}(t) + v_\| h_{xx}(t_r)-
2v_\bot h_{xy}(t_r+T/2) + v_\bot h_{xy}(t_r),
\nonumber
\\
v_{1y}(t) &=& -v_\| h_{xy}(t) + v_\bot h_{yy}(t) - v_\| h_{xy}(t_r)+
2v_\| h_{xy}(t_r+T/2) - v_\bot h_{yy}(t_r).
\label{geosolutions-second}
\end{eqnarray}

\subsubsection{The phase shift}

We now calculate the action for the $r$th atom as it travels through
along the upper path through the interferometer. To
first order in $h_{ij}$,
\begin{eqnarray}
S_{geo}^{cl}[\vec{x}(t_r),\vec{x}(t_r+T);\gamma_{+1}] &\equiv&
\frac{1}{2}m\int_{t_r}^{t_r+T} \left\{[\vec{v}_0(t)+\vec{v}_1(t)]^2
+ v_0^iv_0^jh_{ij}\right\} dt,
\nonumber
\\
&\approx&
\frac{1}{2}m\int_{t_r}^{t_r+T} \vec{v}_0^{\>2}\> dt +
m\int_{t_r}^{t_r+T}\left\{\vec{v}_0\cdot\vec{v}_1(t)
+ \frac{1}{2}v_0^iv_0^jh_{ij}\right\}dt,
\label{ActionGeo}
\end{eqnarray}
where the superscript ``cl'' stands for ``classical'', and denotes
that the integration is over a classical path through
the interferometer. The $\vec{v}_0^{\>2}$ term in
Eq.~$(\ref{ActionGeo})$ is the same for
both paths through the interferometer, and will not contribute to
the phase shift. It can be neglected. The last term $v_0^iv_0^jh_{ij}$
can be trivially integrated using Eqs.~$(\ref{firsthalf})$ and
$(\ref{secondhalf})$, while the $\vec{v}_0\cdot\vec{v}_1$ term can be
integrated using Eqs.~$(\ref{geosolutions-first})$ and
$(\ref{geosolutions-second})$. We then get
\begin{align}
\tilde{S}^{geo}_{cl}[\vec{x}(t_r),\vec{x}(t_r+T);\gamma_{+1}] =&
T\left[v_\|^2h_{xx}(t_r)+v_\bot^2h_{yy}(t_r)\right]
\nonumber
\\
&-
\frac{1}{2}\int_{t_r}^{t_r+T}
\bigg[v_\|^2h_{xx}(t)+v_\bot^2h_{yy}(t)\bigg]dt
\nonumber
\\
&+
2 v_\bot v_\|T\bigg[h_{xy}(t_r) -h_{xy}(t_r+T/2)\bigg]
\nonumber
\\
&-
v_\bot v_\|\left(\int_{t_r}^{t_r+T/2} h_{xy}(t)\>dt -
\int_{t_r+T/2}^{t_r+T} h_{xy}(t)\> dt \right),
\label{ActionGeoFinal}
\end{align}
where the tilde denotes the fact that we have dropped the
$\vec{v}^{\>2}$ term from the action. We get the action
$\tilde{S}_{geo}^{cl}[\vec{x}(t_r),\vec{x}(t_r+T);\gamma_{-1}]$ for
the atom if it took the lower path through the interferometer by
taking $v_\bot\to -v_\bot$ in
$\tilde{S}_{geo}^{cl}[\vec{x}(t_r),\vec{x}(t_r+T);\gamma_{+1}]$. The
first two terms in Eq.~$(\ref{ActionGeoFinal})$ do not change, while
the last two terms flip sign. Thus, from
Eq.~$(\ref{def-phase-shift})$, the phase shift of the atom is
\begin{align}
\Delta\phi_{geo}(t_r) = -\frac{2m}{\hbar} L_\|L_\bot
&\Bigg\{2\left[\frac{h_{xy}(t_r+T/2) -h_{xy}(t_r)}{T}\right]
\nonumber
\\
&+
\frac{1}{T^2} \left(\int_{t_r}^{t_r+T/2} h_{xy}(t)\>dt -
\int_{t_r+T/2}^{t_r+T} h_{xy}(t)\>dt \right)\Bigg\}.
\label{PhaseGeo}
\end{align}
After taking the Fourier transform,
\begin{equation}
h_{xy}(t) \equiv \int_{-\infty}^\infty \frac{d\omega}{2\pi}
h_{\times}(\omega) e^{-i\omega t}, \qquad \Delta\phi_{geo} (t) \equiv
  \int_{-\infty}^\infty \frac{d\omega}{2\pi} \Delta\phi_{geo}(\omega)
  e^{-i\omega t},
\label{Fourier}
\end{equation}
the calculated phase shift for Fig.~$\ref{H-V-MIGO}$ using TT
coordinates simplifies to
\begin{align}
\Delta\phi_{geo}(\omega) \equiv &-\frac{m}{2\hbar}L_\| L_\bot i\omega
h_{\times}(\omega) e^{-i\omega T/2} F_{geo}(\omega T),
\nonumber \\
F_{geo}(\omega T) = & \bigg[\hbox{sinc}\,(\omega
T/4)\bigg]^2-4e^{i\omega T/4} \hbox{sinc}\,(\omega T/4),
\label{PhaseGeoRes}
\end{align}
where $\hbox{sinc}\,(x) = \sin(x)/x$. The first result of our comparison of
phase-shift calculations for the interferometer shown in
Fig.~$\ref{H-V-MIGO}$ is given in Eq.~$(\ref{PhaseGeoRes})$. The
derivation of these equations is based on the geodesic equation of
motion using the TT coordinates. We will find that the phase shift
calculated in all three approaches in this paper will have this same
overall form, and will differ only in the resonance function $F_{geo}$.

\subsection{Geodesic-deviation-EOM-based calculation
of the phase shift}

Reference frames centered on the worldline of an observer, such as the
proper reference frame and general laboratory frame, are less commonly
used than the TT coordinates used in the previous subsection. We
therefore begin the calculation of the phase shift for
Fig.~$\ref{H-V-MIGO}$ in the proper reference frame with a brief
review of this frame.

\subsubsection{The proper reference frame}

How the proper reference frame is constructed is described as follows:
\begin{quote}
``Consider an observer (freely falling or accelerated, it doesn't
matter so long as the acceleration is slowly varying). Let the
observer carry with herself a small Cartesian latticework of
measuring rods and synchronized clocks (a `proper reference frame'
in the sense of Sec.~13.6 of MTW, with spatial coordinates
$x^j$ that measure proper distance along orthogonal axes).'' (From
page 339 of \cite{Thorne}.)
\end{quote}
In this frame the coordinate $x^i(t)$ is the position of a
nonrelativistic test particle travelling along its worldline
\textit{relative} to the worldline of the observer at her
propertime $t$. The magnitude of $x^i(t)$ is then the distance
between the worldline of the test particle, and the worldline of
the observer at the same instant of time $t$. This frame is only
valid when the test particle is moving nonrelativistically, and
the size of the apparatus is small compared to the wavelength of the
gravitational wave. The velocity $v^i(t)=\dot{x}^i$ is then the
velocity of the test particle measured \textit{relative} to the
observer.

In the absence of all other forces, both the test particle and the
observer travel along geodesics in the spacetime, and thus their
motion are each, separately, governed by geodesic EOMs. By
subtracting the geodesic EOM of the observer from that of the test
particle, and then expanding the resultant to first order in
$x^i(t)$, the motion of the test particle \textit{as seen by the
observer} is governed by
\begin{equation}
\frac{d^2x^i}{dt^2} = - R^i_{0j0}x^j.
\label{geodeviation}
\end{equation}
This is the geodesic \textit{deviation} EOM, and it holds as long
as the local Riemann curvature tensor of the spacetime $R^i_{0j0}$
varies so slowly with $x^i$ that it can be approximated as a
constant. This condition holds in the long-wavelength
limit of the gravitational wave.

As observed in \cite{Thorne} for LIGO (the \textit{L}aser
\textit{I}nterferometer \textit{G}ravitational-wave
\textit{O}bservatory), if the characteristic size of the
interferometer $L$ is much smaller than the reduced wavelength
$\lambdabar$ of the gravitational wave the interferometer is contained
entirely within the proper reference frame. One does not need to use TT
coordinates. Like LIGO, the characteristic size of
the interferometers considered in \cite{nmMIGO, CrysMIGO} is also
smaller than the reduced wavelength of the gravitational wave
\cite{nmMIGO, CrysMIGO}, and thus the use of the proper reference
frame was also valid.  We shall assume that this is true for the
interferometer shown in Fig.~$\ref{H-V-MIGO}$ as well.

We will also follow Thorne \cite{Thorne}, and take
$R^i_{0j0}=-\ddot{h}^i_j/2$. This relation between $R^i_{0j0}$ and
$h^i_j$ holds only when the ``TT gauge'' is taken. Because of the
naturalness of its construction, it should be expected that the TT
gauge\textemdash where the number of independent components of
$h_{ij}$ equals the number of degrees of freedom for the
gravitational wave\textemdash can be taken for the proper reference
frame as well as for the TT coordinates,
but this has not been shown explicitly in the literature
\cite{Comment}. As we shall see in \textbf{appendix B}, this can be
done explicitly in the general laboratory frame. It can thus be done
in the proper reference frame since it was shown in \cite{GLF} that
the general laboratory frame reduces to the proper reference frame in
the long-wavelength limit for the gravitational waves.

\subsubsection{Classical Dynamics in the proper reference frame}

For our calculation of the phase shift in the proper reference frame
using the geodesic deviation EOM, the observer is fixed on the
initial beam splitter, and the motion of all objects, and thus the
phase shift, is measured with respect to it (see
Fig.~$\ref{H-V-MIGO}$). In this reference frame, the action for a test
particle is (see \cite{GLF} or \cite{ADS1995} for a derivation)
\begin{equation}
S_{GD} = m \int \left( \frac{1}{2}\vec{v}^{\>2} - v_i x^j
\Gamma^i_{0j}\right)dt = \frac{1}{2} m\int \left(\vec{v}^{\>2} - v^i
x^j\dot{h}_{ij}\right)dt,
\label{GDAction}
\end{equation}
where the subscript ``GD'' stands for ``geodesic deviation'', and we have
dropped the $mc^2$ term. As in Eq.~(4) of \cite{Thorne}, we have taken the
gravitational wave to be in the TT gauge in
Eq.~$(\ref{GDAction})$. The resulting Hamiltonian (see also
\cite{Synge1935}) in this case is
\begin{equation}
H_{GD} = \frac{\vec{p}^{\>2}}{2m} + \Gamma^i_{0j}
x^jp_i=\frac{\vec{p}^{\>2}}{2m} + \frac{1}{2}\dot{h}_{ij}x^ip^j,
\label{GDHam}
\end{equation}
to lowest order in $h_{ij}$. It is straightforward to see from
Eq.~$(\ref{GDAction})$ that the EOM for the test particle is the
geodesic \textit{deviation} equation Eq.~$(\ref{geodeviation})$. We
once again solve perturbatively for the classical trajectories,
which are determined now by Eq.~$(\ref{geodeviation})$, and take
$\vec{x}(t)=\vec{x}_0(t)+\vec{x}_1(t)$.

\subsubsection{Boundary conditions and mirror effects}

\textit{Unlike} the geodesic EOM [Eq.~$(\ref{geoeqn})$], in the
geodesic \textit{deviation} EOM [Eq.~$(\ref{geodeviation})$] the
acceleration depends on the \textit{position} of the test particle, and
\textit{not} on its velocity. Consequently, when a gravitational wave passes
through the system, it shifts the position of \textit{all} parts
of the interferometer \textit{except} the initial beam splitter,
where $x^i\equiv 0$ due to our choice of origin. We can thus take
$\vec{v}_1(t_r) =0$ as an initial condition for $\vec{v}_1(t)$ once
again. In addition, from the force lines drawn in
Fig.~$\ref{H-V-MIGO}$ we see by symmetry that it is the $\times$
polarization that contributes to the phase shift, and not the $+$
polarization. For this polarization, the motion induced on the final
beam splitter by the gravitational wave also does not affect the
overall phase shift.

While the motion of the final beam splitter does not affect the phase
shift of the atoms, the motion of the mirrors in response to the
gravitational wave will. This motion changes the boundary condition
for the atom at the mirrors from the simple expression Eq.~$(\ref{bc})$.

Let $\vec{X}$ denote the position of the mirror placed in the upper
path of the interferometer, and let $\vec{V}$ denote its velocity. In
absence of the gravitational wave the mirrors do not move, and we can
take $X = L_\|/2 + X_1$ and $Y = L_\bot/2 + Y_1$, where
$\vec{X}_1$ are fluctuations in the positions of the mirror due to the
gravitational wave. Because of these fluctuations, the boundary condition
Eq.~$(\ref{pertGeobc})$ for $v_{1y}(t)$ at the mirror now becomes
\begin{equation}
v_{1y}(t_r+T^{+}/2) = -v_{1y}(t_r+T^{+}/2) + V_{1y}(t_r+T/2),
\label{GeoDevBC}
\end{equation}
where $V_{y} = \dot{Y}=\dot{Y}_1$.

To determine the response $V_{1y}(t)$ of the mirror to the
gravitational wave, we follow \cite{nmMIGO, CrysMIGO}, and model the
mirror, and its connection to the frame of the interferometer, as a
spring with a quality factor $Q$ and resonance frequency $\omega_{0}$
(see also section 37.6 of \cite{MTW}). This frequency depends on the
size of the interferometer as well as its material properties. From
the equivalence principle, the motion $\vec{X}(t)$ of the mirror, like
that of the atom, is also a solution of Eq.~$(\ref{geodeviation})$,
but because the mirrors are connected to damped springs,
\begin{eqnarray}
\ddot{X_1} + \frac{\omega_0}{Q} \dot{X_1} + \omega_0^2
X_1(t)& =&
\frac{1}{4} L_\|\ddot{h}_{xx} + \frac{1}{4} L_\bot\ddot{h}_{xy},
\nonumber \\
\ddot{Y_1} + \frac{\omega_0}{Q} \dot{Y_1} + \omega_0^2 Y_1(t) &=&
\frac{1}{4} L_\bot\ddot{h}_{yy} + \frac{1}{4} L_\|\ddot{h}_{xy}.
\label{mirrors}
\end{eqnarray}

\subsubsection{The Phase Shift}

The action for the atom along the upper path through the
interferometer in the proper reference frame is
\begin{eqnarray}
S_{GD}^{cl}[\vec{x}(t_r),\vec{x}(t_r+T];\gamma_{+1}) &\equiv&
\frac{1}{2}m\int_{t_r}^{t_r+T} \left\{[\vec{v}_0(t)+\vec{v}_1(t)]^2
- v_0^ix_0^j\dot{h}_{ij}\right\}\>dt
\nonumber
\\
&\approx&
\frac{1}{2}m\int_{t_r}^{t_r+T} \vec{v}_0^2\> dt + m\int_{t_r}^{t_r+T}
\left\{v_{1i} -
\frac{1}{2}x_0^j\dot{h}_{ij}\right\}\frac{dx_{0}^i}{dt}\>dt,
\label{ActionGeoDev}
\end{eqnarray}
where once again ``cl'' stands for ``classical''. The
$\vec{v}_0^{\>2}$ term is once again the same for both paths through
the interferometer, and can be neglected. Performing an integration by
parts, and using Eq.~$(\ref{geodeviation})$, we get
\begin{align}
\frac{1}{m}\tilde{S}^{GD}_{cl}[\vec{x}(t_r),\vec{x}(t_r+T);\gamma_{+1}]
= & x_0^i(t)\left(v_{1i}(t)-\frac{1}{2} x_0^j(t) \dot{h}_{ij}(t)\right)
\Big\vert_{t_r}^{t_r+T/2}
\nonumber \\
&+ x_0^i(t)\left(v_{1i}(t)-\frac{1}{2} x_0^j(t)
\dot{h}_{ij}(t)\right)\Big\vert_{t_r+T/2}^{t_r+T} \nonumber
\\
&+ \frac{1}{2}\int_{t_r}^{t_r+T/2}x_0^iv_0^j \dot{h}_{ij}\>dt+
\frac{1}{2}\int_{t_r+T}^{t_r+T} x_0^iv_0^j \dot{h}_{ij}\>dt.
\label{ActionGeoDev2}
\end{align}
We integrate by parts once again, and then use Eqs.~$(\ref{firsthalf})$,
and $(\ref{secondhalf})$ to get
\begin{align}
\frac{1}{m}\tilde{S}^{GD}_{cl}[\vec{x}(t_r),\vec{x}(t_r+T);\gamma_{+1}]
= & L_\|\left\{ v_{1x}(t_r+T) -\frac{1}{2} L_\|
\dot{h}_{xx}(t_r+T)\right\}
\nonumber \\
&+
\frac{1}{2}L_\bot\left\{v_{1y}(t_r+T^{-}/2)-v_{1y}(t_r+T^{+}/2)\right\}
\nonumber \\
&+
\frac{1}{2}x_0^i(t) v_0^j(t) h_{ij}(t)\Bigg\vert_{t_r}^{t_r+T/2}
+\frac{1}{2}x_0^i(t) v_0^j(t)
h_{ij}(t)\Bigg\vert_{t_r+T/2}^{t_r+T}
\nonumber \\
&-
\frac{1}{2}\int_{t_r}^{t_r+T}v_0^iv_0^j \dot{h}_{ij}\>dt.
\label{ActionGeoDev3}
\end{align}

Obtaining
$\tilde{S}_{GD}^{cl}[\vec{x}(t_r),\vec{x}(t_r+T);\gamma_{-1}]$ the
usual way from
$\tilde{S}_{GD}^{cl}[\vec{x}(t_r),\vec{x}(t_r+T);\gamma_{+1}]$,
we arrive at
\begin{align}
\Delta\phi_{GD}(t_r) =\frac{m}{\hbar} \Bigg\{ &
L_\|\Big[v_{1x}^{(+1)}(t_r+T) -v_{1x}^{(-1)}(t_r+T) -v_\bot
h_{xy}(t_r+T)
\nonumber \\
&+
v_\bot h_{xy}(t_r+T/2)\Big]+
L_\bot\left[v_{1y}^{(+1)}(t_r+T^{-}/2) +v_{1y}^{(-1)}(t_r+T^{-}/2)
\right]
\nonumber \\
&-
\frac{1}{2}L_\bot\left[V_{1y}^{(+1)}(t_r+T/2)
+V_{1y}^{(-1)}(t_r+T/2) \right]
\nonumber \\
&- 2v_\| v_\bot \left(\int_{t_r}^{t_r+T/2} h_{xy}(t)\>dt -
\int_{t_r+T/2}^{t_r+T} h_{xy}(t)\> dt \right)\Bigg\},
\label{GDPhase1}
\end{align}
where we have used the boundary condition Eq.~$(\ref{GeoDevBC})$. The
superscripts $(\pm 1)$ denote whether it is the velocity of
the atom or mirror along the top [represented by the superscript
($+1$)] or the bottom [represented by the superscript $(-1)$] path
through the interferometer.

We note that to lowest order in $h_{ij}$ Eq.~$(\ref{geodeviation})$
becomes
\begin{equation}
\frac{dv_1^i}{dt} = \frac{1}{2}\ddot{h}^i_j x_0^i(t).
\label{FirstGeoDev}
\end{equation}
We then take the component of Eq.~$(\ref{FirstGeoDev})$ along the
$x$ direction for an atom travelling along the lower path, and
subtract it from the same equation for the atom travelling along
the upper path. Since $x_0^{(+1)}(t)=x_0^{(-1)}(t)$, while
$y^{(+1)}_0(t)=-y^{(-1)}_0(t)$, we integrate and find
\begin{equation}
v_{1x}^{(+1)}(t_r+T) - v_{1x}^{(-1)}(t_r+T) =  v_\bot\Big\{h_{xy}(t_r+T)
- 2 h_{xy}(t_r+T/2)+h_{xy}(t_r)\Big\}.
\end{equation}
Similarly, if we take the sum of these equations along the $y$
direction, we get
\begin{equation}
\nonumber \\
v_{1y}^{(+1)}(t_r+T^{-}/2) + v_{1y}^{(-1)}(t_r+T^{-}/2) =
\frac{1}{2} L_\|\dot{h}_{xy}(t_r+T/2) - v_\|\Big\{h_{xy}(t_r+T/2)
- h_{xy}(t_r)\Big\}.
\end{equation}
Consequently,
\begin{align}
\Delta\phi_{GD}(t_r) =&-\frac{2m}{\hbar} L_\|
L_\bot\Bigg\{
\frac{1}{4L_\|}\bigg[V_y^{(+1)}(t_r+T/2)+V_y^{(-1)}(t_r+T/2)\bigg]
\nonumber
\\
&- \frac{1}{4}\dot{h}_{xy}(t_r+T/2)+ \frac{h_{xy}(t_r+T/2) -
h_{xy}(t_r)}{T} \nonumber
\\
&+
\frac{1}{T^2} \left(\int_{t_r}^{t_r+T/2} h_{xy}(t)\>dt -
\int_{t_r+T/2}^{t_r+T}
  h_{xy}(t)\>dt\right)\Bigg\}.
\label{GDPhase2}
\end{align}
Using Eq.~$(\ref{Fourier})$ and the steady-state solution for
Eq.~$(\ref{mirrors})$, then
\begin{equation}
V_{1y}^{(+1)}(t_r+T/2) + V_{1y}^{(+1)}(t_r+T/2) =
-\frac{1}{2}L_\|\int\frac{d\omega}{2\pi}
\frac{i\omega^3}{\omega^2-\omega_0^2+
i\omega\omega_0/Q}e^{-i\omega(t_r+T/2)},
\end{equation}
and we find that
\begin{align}
\Delta\phi_{GD}(\omega) \equiv &-\frac{m}{2\hbar}L_\| L_\bot i\omega
h_{\times}(\omega) e^{-i\omega T/2} F_{GD}(\omega T),
\nonumber \\
F_{GD}(\omega T) = &1-2e^{i\omega T/4} \hbox{sinc}\,(\omega T/4) +
\bigg[\hbox{sinc}\,(\omega T/4)\bigg]^2 - \frac{1}{2}
\frac{\omega^2}{\omega^2-\omega_0^2 + i\omega\omega_0/Q}.
\label{F-GD}
\end{align}
The second result of our comparison of phase-shift calculations for
the interferometer shown in Fig.~$\ref{H-V-MIGO}$ is given in Eq.
$(\ref{F-GD})$. The derivation of these equations is based on the
geodesic deviation equation of motion using the proper reference
frame. The phase has precisely the same form as
Eq.~$(\ref{PhaseGeoRes})$; only the resonance function $F_{GD}$ has
changed.

\begin{table}[ptb]
\begin{center}
\begin{tabular}[c]{|c|c|c|}\hline \textit{Quantity} &
Geodesic-equation based& Geodesic-deviation-equation based\\
\hline \textit{Action} & $m \int dt\left[v^iv^j(\delta_{ij} +
  h_{ij})\right]/2$& $m\int dt \left[\vec{v}^{\>2} - v^i
x^j\dot{h}_{ij}\right]/2$\\
\textit{EOM} & $\ddot{x}^i = -\dot{h}^i_j \dot{x}^j$&
$\ddot{x}^i=\ddot{h}^i_j x^j/2$\\
\textit{Hamiltonian} & $\left(\vec{p}^{\>2} - h_{ij}p^ip^j\right)/2m$&
$\vec{p}^{\>2}/2m + \dot{h}_{ij}x^ip^j/2$\\
\textit{Resonance Function} & $\bigg[\hbox{sinc}\,(\omega
T/4)\bigg]^2 - 4e^{i\omega T/4}
\hbox{sinc}\,(\omega T/4)\>$&$1-2e^{i\omega T/4}
\hbox{sinc}\,(\omega T/4)+ \bigg[\hbox{sinc}\,(\omega T/4)\bigg]^2$\\
 & &$-\omega^2/\left(\omega^2-\omega_0^2 + i\omega\omega_0/Q\right)/2$\\
\textit{Low-frequency Phase Shift}&$-3mL_\| L_\bot
\Gamma^x_{0y}(\omega)/\hbar$&$-mL_\| L_\bot T
R_{0x0y}(\omega)/2\hbar$\\
\hline
\end{tabular}
\caption{A comparison between the geodesic EOM and the geodesic
  deviation EOM descriptions of test-particle dynamics, and the phase
  shifts calculated for Fig.~$\ref{H-V-MIGO}$ using the two EOMs.}
\label{Comparison}
\end{center}
\end{table}

\section{The Low-frequency Limit, and Determining the Correct
Phase Shift Expression}

A summary of the geodesic- and geodesic-deviation-based EOM
approaches to the description of the motion of test particles in
the presence of a gravitational wave, along with the resonance functions for the
phase shifts calculated with both approaches, is listed in Table
$\ref{Comparison}$. It is readily apparent from the form of $F_{geo}$ and
$F_{GD}$ that the corresponding phase shifts $\Delta\phi_{geo}$ and
$\Delta\phi_{GD}$ \textit{do not agree with each other} even
if the Lorentzian term in $F_{GD}$ is neglected. Moreover, as we will
see in the next section, both phase shifts are different from the
phase shift calculated previously in the literature as well. There are
thus \textit{three} different expressions for the phase shift for the
\textit{same} matter-wave interferometer. To determine the correct
expression, we consider the low-frequency limit of these expressions,
and compare their limiting forms to the phase shift of caused by
a static Riemann curvature tensor.

\subsection{How the Correct Phase Shift is to be Determined}

The motion of test particles in \textit{stationary} spacetimes has
been well studied experimentally both directly,
through measurements of the advance of the perihelion of Mercury, the
deflection of light, and the gravitational redshift \cite{MTW} in
the weak gravity limit, and indirectly, through the slow-down of
the Taylor-Hulse pulsar \cite{Taylor-Hulse} in the strong gravity
limit. It has also been well studied theoretically through analyses of
various black hole geometries \cite{Chandra}. The motion of test
particles in \textit{nonstationary} spacetimes, such as when a
gravitational wave is present, has not been established to the same level of
certainty. On an experimental level, until LIGO has detected a
gravitational wave, there will have been no \textit{direct}
measurements of the motion of test particles in nonstationary
spacetimes. On a theoretical level, while it would \textit{seem} that
the motion of test particles in nonstationary spacetimes has been well
studied, a recent analysis \cite{GLF} based on the general laboratory
frame has detailed differences in the dynamics of test particles
in nonstationary versus stationary spacetimes.

There is, however, is no disagreement\textemdash on either on a
theoretical or an experimental level\textemdash over the motion of
text particles in \textit{stationary} spacetimes. Indeed, it was
explicitly demonstrated that in this case the general laboratory frame
\cite{GLF} (and thus the proper reference frame) goes over to the
standard description of the motion of test particles in stationary
spacetimes using the geodesic EOM.

The phase shift $\Delta\phi_{static}$ of a matter-wave
interferometer due to a \textit{static} curvature is well known,
and can be written as $\Delta\phi_{static} \sim mA^{ij}T
R_{0i0j}^{static}/\hbar$, where $\vert A^{ij}\vert$ is the area of
the interferometer, $T$ is the transit time of the atom through the
interferometer, and $R_{0i0j}^{static}$ are the components of
the static Riemann curvature tensor in the plane of the
interferometer (see \cite{Anan}, Eq.~$(2.6)$ of \cite{GRG}, and
Eq.~(2.23) of \cite{Papini}). While the calculation of the calculation
of $\Delta\phi_{static}$ in \cite{GRG} and \cite{Papini} was based on
general relativity, $\Delta\phi_{static}$ can be derived using
only Newtonian gravity\textemdash where the static curvature is
formed from gradients of $g$\textemdash as was done in \cite{Anan}.
Newtonian gravity has been experimentally verified, of course, and
the effect of $g$ on the phase shift of massive particles in
matter-wave interferometers is well known for neutrons \cite{COW},
and for atoms \cite{ChuKas, PCC}. The effect of the static Riemann
curvature of the Earth has also been seen in measurements of the
\textit{gradients} in $g$ using differential measurements of the phase
shift caused by $g$ at differing heights \cite{PCC, Kas}. While
indirect, these all serve as experimental indications that
$\Delta\phi_{static}$ holds. Indeed, there are indications that the
phase shift induced by the static Riemann curvature of the Earth has
already been seen \cite{PrivateKas}.

Consider now the limit where the period of a gravitational wave is
very long compared to the transit time of the atom through the
interferometer. The gravitational wave does not oscillate
appreciably in the time it takes for the atom to traverse the
interferometer, and to the atom, the curvature of the spacetime
caused by the gravitational wave is effectively a static one.
Correspondingly, the phase shift induced by the gravitational wave
should have the same form as the phase shift induced by a static
Riemann curvature. This correspondence between the gravitational
wave-induced, and the static-curvature-induced phase shifts serves
as a check of our results. When it is combined with the
experimental results \cite{COW, ChuKas, Kas, PCC}, we conclude
that the \textit{correct} phase shift must have a low-frequency
limit that has the same \textit{form} as $\Delta\phi_{static}$.

\subsection{The Low-frequency Limit}

Unlike $F_{geo}$, which depends only on the transit time $T$ of
the atoms through the interferometer, and the period $2\pi/\omega$ of
the gravitational wave, $F_{GD}$ depends on the period $2\pi/\omega_0$
of the resonance frequency of the mirror-interferometer assembly as
well. Both $T$ and $\omega_0$ are fixed by the design of the
interferometer, while $\omega$ can, in principle, take a broad range
of values. We note, however, that the resonance frequency of the
interferometer can be estimated as $\omega_0\sim 2\pi v_{sound}/L_\|$,
where $v_{sound}$ is the speed of sound of the material used to
construct the interferometer. In almost all cases, $v_\|<<v_{sound}$,
and since $1/T=L_\|/v_\|$, $2\pi/\omega_0<< T$.

If the transit time of
the atom is much shorter than the period of the gravitational wave so
that $T<<2\pi/\omega$, then $\omega<<\omega_0$ as well. In this
low-frequency limit, the gravitational wave will not oscillate in the
time it takes the atom to traverse the interferometer, nor will the
mirrors oscillate appreciably during this time. This, then,
corresponds to the static-limit conditions considered above.

From Eq.~$(\ref{PhaseGeoRes})$, we see that in this low-frequency
limit where $\omega T/4 << \pi/2$, $F_{geo}\approx -3$. Then
\begin{equation}
\Delta\phi_{geo}(\omega) \approx \frac{3m}{2\hbar} L_\| L_\bot i\omega
h_{\times}(\omega) = -3\frac{m}{\hbar} L_\| L_\bot \Gamma^x_{0y}(\omega),
\label{PS-Geo-LF}
\end{equation}
since $\Gamma^x_{0y}(\omega) = -i\omega \dot{h}_\times(\omega)/2$. Thus,
$\Delta\phi_{geo}(\omega)$ is proportional to the Fourier transform of the
\textit{Christoffel symbols} for the gravitational wave in the low
frequency limit, and \textit{not} the Riemann curvature tensor. It
thus has an incorrect limiting form.

If we now take the low-frequency limit of Eq.~$(\ref{F-GD})$, we
find that $F_{GD}\approx -i\omega T/2$ to lowest order, so that
\begin{equation}
\Delta\phi_{GD}(\omega) \approx -\frac{m}{4\hbar} L_\|
L_\bot T\omega^2 h_{\times}(\omega) = -\frac{m}{2\hbar} L_\| L_\bot
R_{0x0y}(\omega),
\label{PS-GD-LF}
\end{equation}
since $R_{0x0y}(\omega) = \omega^2 h_{\times}(\omega)/2$. Thus,
unlike $\Delta\phi_{geo}(\omega)$, $\Delta\phi_{GD}(\omega)$
\textit{is} proportional to the Riemann curvature tensor, and
\textit{has the correct limiting form}.

\begin{figure}[ptb]
\begin{center}
  \includegraphics[width=1.0\textwidth]{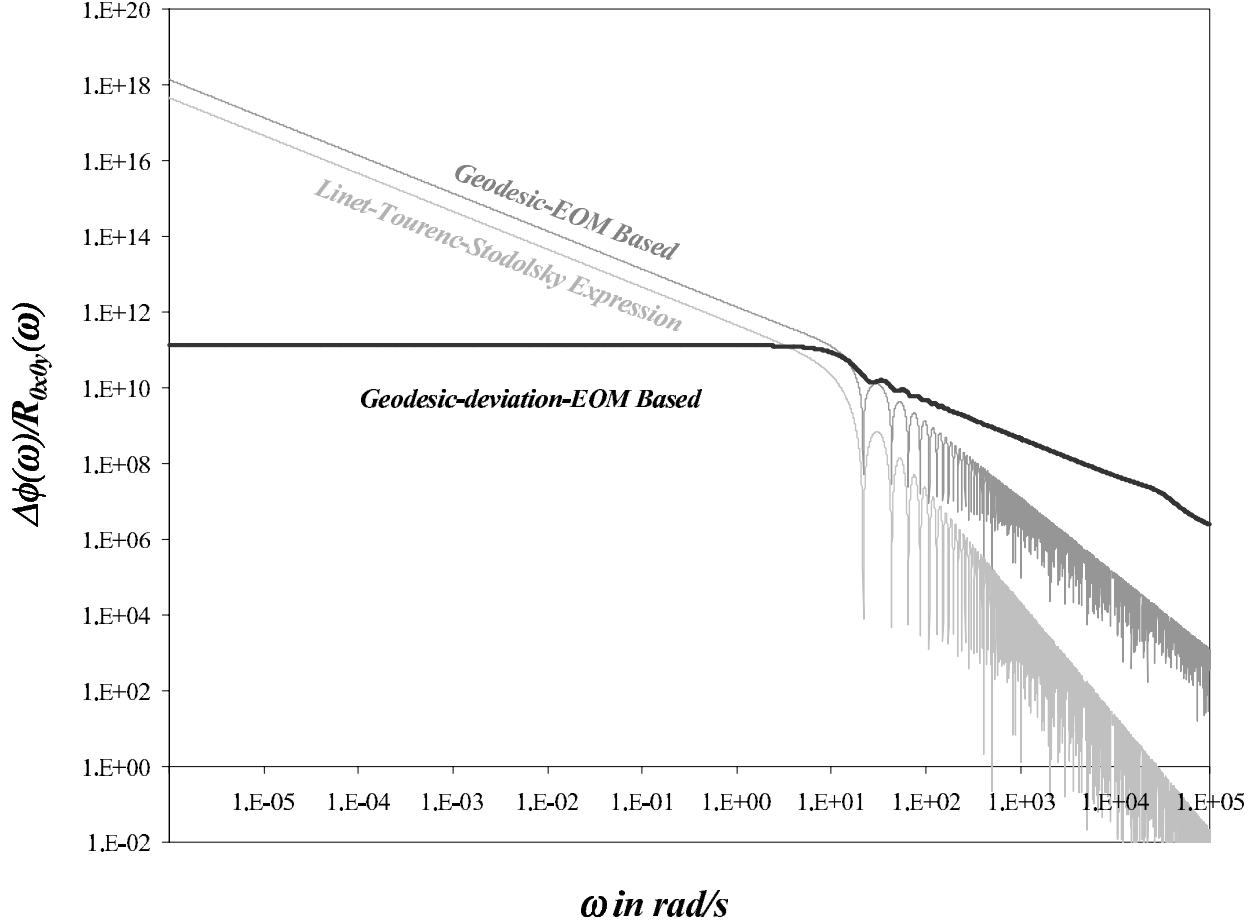}
\end{center}
\caption{A plot of $\Delta\phi(\omega)/R_{0x0y}(\omega)$ versus
$\omega$ for $\Delta\phi_{GD}$, $\Delta\phi_{geo}$, and
$\Delta\phi_{LTS}$, for the interferometer Fig.~$\ref{H-V-MIGO}$ using
$L_\bot = 1.2$ m, $L_\| = 4$ km, $T=0.58$ s, $Q=1$, $\omega_0 =
31,000$ rad/s, and used $^6$Li with $m=1\times 10^{-23}$ gm. }
\label{phase-shift}
\end{figure}

The differences in the low-frequency behavior of
$\Delta\phi_{geo}$ and $\Delta\phi_{GD}$ are shown
Fig.~$\ref{phase-shift}$, where we have graphed
$\Delta\phi(\omega)/R_{0x0y}(\omega)$ versus $\omega$ for both
expressions of the phase shift. For completeness, we have also
included in this comparison the phase shift $\Delta\phi_{LTS}$ for the
matter-wave interferometer Fig.~$\ref{H-V-MIGO}$ calculated using the
Linet-Tourenc-Stodolsky formula \cite{Linet-Tourrenc, Stodolsky}. As
we mentioned in the introduction, a third calculation of the phase
shift of Fig.~$\ref{H-V-MIGO}$ based also on the geodesic EOM is
necessary, and in the next section this phase shift $\Delta\phi_{LTS}$ is
explicitly calculated for our interferometer using the
Liner-Tourrenc-Stodolsky formula. For now, we note that the graphs in
Fig.~$\ref{phase-shift}$ were plotted using the parameters of the
``Earth-base horizontal MIGO'' described \cite{nmMIGO}, which has
the same configuration as the interferometer we consider here in
Fig.~$\ref{H-V-MIGO}$. This interferometer has a $L_\bot = 1.2$ m,
$L_\| = 4$ km, $T=0.58$ s, $Q=1$, $\omega_0 = 31,000$ rad/s, and
used $^6$Li with $m=1\times 10^{-24}$ gm. The plots in
Fig.~$\ref{phase-shift}$ cover the same frequency range considered
in \cite{nmMIGO}. Notice that for frequencies much less than
$2\pi/T = 11$ rad/s, $\Delta\phi_{GD}/R_{0x0y}$
approaches a constant, indicating that
$\Delta\phi_{GD}\propto R_{0x0y}$ for low frequencies,
as expected, while both $\Delta\phi_{geo}/R_{0x0y}$, and
$\Delta\phi_{LTS}/R_{0x0y}$ do not.

The behavior of all three phase shifts also differ dramatically at
high frequencies. We will consider the high-frequency behavior
of $\Delta\phi_{GD}$ in \textbf{appendix C}.

\section{Previous Phase-shift Calculations}

We now compare results of the above calculations with those
done previously in the literature.

\subsection{Calculations done at the quantum-mechanical level}

Calculations of the phase shift by Linet and Tourrenc
\cite{Linet-Tourrenc}, and Stodolsky \cite{Stodolsky} were both done at the
quantum-mechanical level, but unlike us, they presented a completely
general analysis, and did not take the nonrelativistic limit for the
interfering particles they considered. Their expressions for the phase
shift were thus manifestly covariant. Linet and Tourrenc did not explicitly
define the coordinate system that they used, while Stodolsky only
referred to the presentation by Landau and Lifschitz \cite{L-L} on
synchronous reference frames. It is clear that since a gravitational
wave in the TT gauge has $h_{0\mu}=0$, the TT coordinates is a
synchronous reference frame as well. Since both calculations were
ultimately based on the action for the \textit{geodesic} EOM, they
were implicitly done in the TT coordinates.

Linet and Tourrenc used the WKB approximation in the quasi-classical
limit, and replaced the mass-shall constraint $g^{\mu\nu}p_\mu
p_\nu = -m c^2$ (note that the signature of their metric is
different from ours) with a differential equation in the phase of
the wavefunction for the test particle by making the substitution
$p_\mu = (i/\hbar) \partial_\mu$. Stodolsky worked directly with
the action, and used the quasi-classical approximation we followed
here to calculate the phase shift. The expression for
the phase shift derived by both nevertheless have the same overall
form. The phase shift (given in Eq.~[3.2.3] of \cite{Linet-Tourrenc})
derived by Linet and Tourrenc does, however, differ from the original
phase shift (given in Eq.~(2.4) of \cite{Stodolsky}) derived by
Stodolsky by a factor of 2. Since Stodolsky uses the Linet and
Tourrenc phase shift expression (the equation above Eq.~(6.1) of
\cite{Stodolsky}) to calculate the phase shift for a model
interferometer, we will do so also. In the nonrelativistic limit of
the atoms that we are interested in, this phase shift reduces to
\begin{equation}
\Delta\phi_{LTS} = \frac{1}{2}\frac{m}{\hbar}\oint v^i v^j h_{ij} dt,
\label{Stod-PS}
\end{equation}
which is different from $\Delta\phi_{geo}$, even though both were
calculated in TT coordinates.

To understand the origin of the difference between $\Delta\phi_{LTS}$
and $\Delta\phi_{geo}$, let us write $S_{geo} = S_{geo}^{kin} +
S_{geo}^{pot}$, where
\begin{equation}
S_{geo}^{kin} = \frac{1}{2} m\int \vec{v}^{\>2} dt, \qquad S_{geo}^{pot} =
\frac{1}{2} m\int v^iv^j h_{ij}dt,
\label{factorize}
\end{equation}
and the superscript ``kin'' stands for ``kinetic'', while the
superscript ``pot''stands for ``potential''. Clearly,
$\Delta\phi_{geo} = \Delta\phi^{kin}_{geo} +
\Delta\phi^{pot}_{geo}$. Since $\Delta\phi_{LTS} =
\Delta\phi^{pot}_{geo}$, contributions to the total phase shift
$\Delta\phi^{kin}_{geo}$ from changes to the kinetic energy of the
test particle as it traverses the interferometer was not taken into
account in Eq.~$(\ref{Stod-PS})$. This kinetic term cannot be neglected,
however. The passage of the gravitational wave through the system will
cause shifts in the velocity of the atom as it traverses the
interferometer, and thus to its kinetic energy. In fact, by using
Eqs.~$(\ref{geosolutions-first})$ and $(\ref{geosolutions-second})$,
it is straightforward to show that $\Delta\phi^{kin}_{geo} =
-2\Delta\phi^{pot}_{geo} + B.T.$, where $B.T.$ are boundary terms for the
atoms at the mirrors. These terms do not vanish, but instead
contribute the $4\>\hbox{sinc}\>(\omega T/4)$ term to $F_{geo}$.

If we nonetheless use $\Delta\phi_{LTS}$ to calculate
the phase shift for Fig.~$\ref{H-V-MIGO}$, we once again
find that $\Delta\phi_{LTS}$ has the same form as
Eqs.~$(\ref{PhaseGeoRes})$ and $(\ref{F-GD})$,
\begin{align}
\Delta\phi_{LTS}(\omega) \equiv &-\frac{m}{2\hbar}L_\| L_\bot i\omega
h_{\times}(\omega) e^{-i\omega T/2} F_{LTS}(\omega T),
\nonumber \\
F_{LTS}(\omega T) = &-\bigg[\hbox{sinc}\,(\omega T/4)\bigg]^2.
\label{F-Stod}
\end{align}
The final result of our comparison of phase-shift calculations for
the interferometer shown in Fig.~$\ref{H-V-MIGO}$ is given in
Eq.~$(\ref{F-Stod})$. While these equations, like
Eq.~$(\ref{PhaseGeoRes})$, are also based on the geodesic equation
of motion using the TT coordinates, $F_{LTS}$ nonetheless
differs from $F_{geo}$ by the term $4\>\hbox{sinc}\,(\omega T/4)$.
This term comes directly from the boundary terms neglected in
Eq.~$(\ref{Stod-PS})$. (When expressed in terms of
resonance functions, Eq.~$(\ref{F-Stod})$ agrees, after the
nonrelativistic limit is taken, with the phase shift calculated in
\cite{Stodolsky} up to overall factors.)

Taking the low-frequency limit, we find that $F_{LTS}\approx -1$.
Thus, like $\Delta\phi_{geo}$, $\Delta\phi_{LTS}$ has the
incorrect low-frequency limiting form. It also has a
\textit{high}-frequency behavior that is dramatically different
from either $F_{geo}$ or $F_{GD}$, as can be seen in
Fig.~$\ref{phase-shift}$. In this limit, while $F_{geo}\sim
1/\omega T$ and $F_{GD} \sim 1$ (see \textbf{appendix C}),
$F_{LTS} \sim 1/(\omega T)^2$, and dies off much more rapidly. As
pointed out in \cite{nmMIGO}, this rapid die off is due to three
effects. Firstly, the $\Delta\phi_{LTS}$ does not include the
boundary terms at the mirror. These terms depend on the value of the
velocity of the atom at the surface of the mirror, and the length
$L_\|/2$, and they mitigate this rapid decrease for
$\Delta\phi_{geo}$. Secondly, $\Delta\phi_{LTS}$ depends on $v^2$,
and is constant as the atom travels through the interferometer,
while $\Delta\phi_{GD}$ depends on the product $x v$,
and increases linearly with the distance the atom travels from the
initial beam splitter. Thirdly, since $\Delta\phi_{LTS}$ depends
only on $v^iv^jh_{ij}$, discontinuous jumps in the velocity due to
the hard wall mirrors were not taken into account in
$\Delta\phi_{LTS}$. From \textbf{appendix C}, these hard-wall jumps
contribute directly to the high-frequency behavior of
$\Delta\phi_{GD}$.

\subsection{Calculations done at the quantum-field-theory level}

\subsubsection{For Spin 0 Fields}

Calculations of the phase shift at the quantum-field-therapy
level were first done by Cai and Papini \cite{Papini}. Their starting
point is the covariant version of the Ginzburg-Landau equation,
but for a charged particle. This is a relativistic Klein-Gordon
equation for a bosonic field $\varphi$ that interacts with itself through
a $\vert\varphi\vert^4$ interaction (Eq.~(2.1) of \cite{Papini}).
By turning off the $\vert\varphi\vert^4$ interaction, and setting
the charge of the particle to zero, one recovers Linet and Tourrenc's
starting point. If one extracts the single-particle Hamiltonian
from this equation, and take the correspondence principle limit of
the EOM, one recovers the geodesic EOM used by Stodolsky. Thus,
like Linet and Tourrenc, and Stodolsky, Cai and Papini work in TT
coordinates.

By acting on $\varphi$ with a phase operator $\chi$, Cai and Papini
was able to solve for a general expression for the phase shift due to
gravitational effects in the weak-gravity limit through a
stationary-phase type of analysis. This operator has the form
[Eq.~(3.1) of \cite{Papini}],
\begin{equation}
\chi=\left\{-\frac{1}{4}\oint_\gamma
dz^\lambda\left(\Gamma_{\alpha,\lambda\beta}
-\Gamma_{\beta,\lambda\alpha}\right)J^{\alpha\beta}
+\frac{1}{2}\oint_\gamma dz^\lambda g_{\lambda\beta}P^\beta\right\},
\label{chi}
\end{equation}
(using our notation) for a closed path where $J_{\alpha\beta}$ is the
generator of Lorentz transformations, and $P^\alpha$ is the generator
of translations. We have dropped the vector potential term since the
atoms we consider in this paper are not charged. As they mentioned,
using Stokes' theorem the first term in $\chi$ can be written as
\begin{equation}
-\frac{1}{4} \int_{\cal D} R_{\mu\nu\alpha\beta}
 J^{\alpha\beta}d\tau^{\mu\nu},
\end{equation}
in the weak-gravity limit; the integral is over the timeline
surface $\cal D$ bounded by $\gamma$, and $d\tau^{\mu\nu}$ is a
2-form on $\cal D$. This expression has been explicitly verified
in \cite{GRG} for the phase shift caused by Riemann curvature
tensor for a \textit{stationary} spacetime.

The second term agrees with Stodolsky's expression for the phase
shift. Indeed, Cai and Papini uses Eq.~$(\ref{chi})$ to calculate the
phase shift of a model interferometer due to the passage of a
gravitational wave [Eq.~(3.15) of \cite{Papini}], and they found that
up to terms of order $(v/c)^2$, the following holds:
\begin{quote}
``The $\Delta\chi_2$ terms in (3.15) agrees with Stodolsky's result
[11] when a misprint in the latter is corrected by inserting an
overall factor $\frac{1}{2}$.'' (From page 415 of \cite{Papini}.)
\end{quote}
The reference [11] cited above is the work by Stodolsky cited as
\cite{Stodolsky} in this paper. Consequently, their expression for the
phase shift, like Linet and Tourrenc's, and Stodolsky's, also has the
incorrect low-frequency limiting form. The effects of the mirrors on
the test particle's phase was not considered by them either.

\subsubsection{For Spin 1/2 Fields}

Bord\'e and coworkers \cite{Borde1, Borde2, Borde3} considered the
use of fermionic atoms in matter-wave interferometers to
measure various gravitational effects, including the detection of
gravitational waves. Their starting point was the Dirac equation for a
two-level, spin 1/2 atom expressed in curved spacetimes using the
tetrad formalism. Similar to Cai and Papini, theirs was a
Hamiltonian-based approach, and they showed that the dominant
contribution by the gravitational wave to the Hamiltonian is from a
term with the form $p_ip_j h^{ij}/2m$, where $m$ is the mass of the
atom, and $p_i$ its momentum [see Eq.~(10) of \cite{Borde1}]. The other
terms in the Hamiltonian due to the gravitational wave are
from very small spin-gravitation couplings. These terms can
be neglected here since we are dealing with atoms without spin.
When this is done, the Hamiltonian they calculated has the same form as
Eq.~$(\ref{geoham})$ for the geodesic EOM, and is manifestly different
from the Hamiltonian Eq.~$(\ref{GDHam})$ for the geodesic
\textit{deviation} EOM. Thus they, too, worked in the TT coordinates.

Although an expression for the phase shift was not calculated in
\cite{Borde1}, they concluded the following:
\begin{quote}
``The phase shift, that one can calculate from the first terms of
Eq.~(10), increases with the velocity of the atoms in contrast with
what happens in the inertial field case and a gravitational wave
detector using atomic interferometry should therefore use relativistic
atoms. In this case, it can be shown directly from Eq.~(5) that the
phase shift is proportional to $h_{(\times,+)}\oint ds/\lambda$ where
$\lambda$ is the atomic de Broglie wavelength, result which is similar
to the optical case.'' (From page 161 of \cite{Borde1}.)
\end{quote}
In a subsequent paper \cite{Borde3}, Bord\'e gave an explicit
expression for the phase shift [see Eq.~(98) of \cite{Borde3}]. It
has, for its spin 0 component, the same expression,
$\Delta\phi_{LTS}$, for its phase shift as Stodolsky's. Thus,
like Stodolsky's approach, any phase shift calculated using this
analysis would have the incorrect low-frequency limiting form as well.

\subsubsection{More General Analysis}

Finally, all of the analyses that we outlined in this paper were done
in the weak-gravity limit. In \cite{Alsing}, a derivation of the
phase shift of a quantum mechanical test particle was done without
making this approximation. Their analysis was also based on
the geodesic-EOM approach, and they arrived same expression for
the phase shift that Stodolsky does for spin 1/2 test particle
with no approximations, while it holds only the first
order in $\hbar$ for spin 0 and spin 1 particles. Consequently, phase
shifts calculated with their expression will also have the same
incorrect low-frequency limiting form as Stodolsky's.

\section{Concluding Remarks}

There has been a preference among those who study matter-wave
interferometry and gravitational waves to use TT coordinates to
analyze the phase shift of an interferometer induced by
gravitational waves instead of the proper reference frame. The
expectation is that irrespective of which reference frame is used to
calculate the phase shift, the final expression will be the same
\cite{Thorne, Unruh-Weiss}. We have shown in this paper that for
matter-wave interferometry \textit{this expectation is wrong}. By
presenting a calculation of the phase shift using the TT coordinates
following the line of the analysis found in the literature
\cite{Linet-Tourrenc, Stodolsky, Papini, Borde1, Borde2, Borde3,
Alsing}, and also a calculation of the phase shift using the proper
reference frame following the analysis found in \cite{nmMIGO,
CrysMIGO}, we have shown that the expressions for the phase shift
obtained using the TT-coordinates \textit{is} different from that
obtained using the proper reference frame. Moreover, we have shown
that the commonly accepted expression for the phase shift
Eq.~$(\ref{Stod-PS})$\textemdash which is the end result of all previous
derivations of the phase shift in the literature \cite{Linet-Tourrenc,
Stodolsky, Papini, Borde1, Borde2, Borde3, Alsing}\textemdash
\textit{is incomplete}, and gives yet a different expression for the
phase shift of Fig.~$\ref{H-V-MIGO}$. This happens even though
derivations of this expression were all done in TT coordinates.

The focus of this paper has been on demonstrating in detail that
differences exist between the various approaches to calculating
the phase shift, and showing explicitly that different expressions
for the phase shift are obtained. To determine which one, if any,
is \textit{correct}, we noted that in the limit of sufficiently
low frequency gravitational waves, the correct expression for the
phase shift \textit{must} reduce to the well known expression for
the phase shift induced by a static curvature. It must therefore
be proportional to components of the Riemann curvature tensor.
This static phase shift can also be calculated within Newtonian
gravity, where the curvature is formed from gradients of the
acceleration due to gravity $g$. The phase shift due to $g$ has
been measured \cite{COW, ChuKas}, and recently has been used to
measure gradients in $g$ as well \cite{PCC, Kas, PrivateKas}. In
addition, there are indications that the phase shift induced by
the static Riemann curvature tensor has already been seen
\cite{PrivateKas}. We have explicitly shown that when the phase shift
is calculated using the proper reference frame and geodesic deviation
EOM, it has the expected limiting form in the low-frequency limit, and
is therefore correct. The other two, both of which used the TT
coordinates and the geodesic EOM, do not, and are thus incorrect.

The implications of this conclusion are immense. In \cite{nmMIGO,
CrysMIGO}, MIGO, the \textit{M}atter-wave \textit{I}nterferometric
\textit{G}ravitational-wave \textit{O}bservatory, was proposed. The
theoretical sensitivity predicted for MIGO was based on a calculation
of its phase shift using the proper reference frame and the geodesic
deviation EOM. The conclusion arrived at in these papers was that by
using \textit{slow} atoms, it would be possible to construct gravitational wave
observatories that are as sensitive as LIGO or LISA (the \textit{L}aser
\textit{I}nterferometer \textit{S}pace \textit{A}ntenna), but would be
orders of magnitude smaller, and have a much broader frequency
response. This conclusion is diametrically opposite from the
conclusions arrived previous in the literature that \textit{fast}
atoms\textemdash indeed, ultrarelativistic ones\textemdash should be
used in the matter-wave interferometer instead \cite{Borde1}.

The conclusion that relativistic atoms should be used in the
detection of gravitational waves with matter-wave interferometers is
the basis on which most of the atomic, molecular and optical physics
community judges the suitability of using
matter-wave interferometers as gravitational wave detectors. The
physical origin of this conclusion by Bord\'e and his
coworkers is the dependence of the phase shift $\Delta\phi_{LTS}$ on
the \textit{square} of the atom's velocity in the geodesic-EOM-based
approach; clearly, the faster the atom, the larger the phase shift
will be. Even though our analysis of the phase shift was done for
nonrelativistic atoms, we note that if the size of the interferometer
is held constant, and only the velocity of the atom is
increased, then in the high-velocity limit, $T\to0$, and eventually $\omega
T<<\pi/4$ for the frequency of gravitational waves expected from
astrophysical sources. The high-velocity limit is thus equivalent to
the low-frequency limit for the gravitational waves, and we have shown
that the low-frequency limit of $\Delta\phi_{LTS}$ is
incorrect. Thus, Bord\'e and coworker's assertion about relativistic
atoms being more suitable for the detection of gravitational waves is
also incorrect.

All previous calculations and estimates of the phase shift in the
literature also neglected the effect the mirrors of the
interferometer have on the phase shift. Indeed, from \textbf{appendix
C} we see that it is precisely because hard wall mirrors are
used\textemdash along with the \textit{instantaneous} impulsive force
that they exert on the atoms when they reflect off\textemdash that the
use of slow-velocity atoms leads to a sensitivity that \textit{increases}
with frequency is obtained. This dependence on frequency is the exact
opposite for LIGO and LISA \cite{Thorne}, and is part of the reason
that MIGO can be so much smaller than either one.

We have not, in this paper, presented the underlying physical
reasons \textit{why} the current TT coordinates approach to
calculating the phase shift fails, and the proper reference frame
succeeds. We shall address this issue in detail later
\cite{ADS-Chiao-2004}. Here, we only note that one may try to modify
the boundary conditions for the atom on the mirrors so that the
phase shift calculated for Fig.~$\ref{H-V-MIGO}$ in TT coordinates
will have the correct low-frequency limit. Doing this is especially
tempting in light of the following statement by Thorne:
\begin{quote}
``However, errors are much more likely in the TT analysis than in
the proper-reference-frame analysis, because our physical
intuition about how experimental apparatus behaves is
proper-reference-frame based rather than TT-coordinate based. As
an example, we intuitively assume that if a microwave cavity is
rigid, its walls will reside at fixed coordinate locations $x^j$. This
remains true in the detector's proper reference frame (aside from
fractional changes of order $(L^2/\lambdabar^2)h$, which are truly
negligible if the detector is small and which the
proper-reference-frame analysis ignores). But it is not true in TT
coordinates; there the coordinate locations of a rigid wall are
disturbed by fractional amounts of order $h$, which are crucial in
analyses of microwave-cavity-based gravity wave detectors.'' (From
page 400 of \cite{Thorne}.)
\end{quote}
We find that while it may be possible to suitably modify the
boundary conditions to bring $\Delta\phi_{geo}$ in line with the
low frequency limit, doing so will be \textit{ad hoc}, and will
not explain the underlying physical reason for the differences
between $\Delta\phi_{geo}$ and $\Delta\phi_{GD}$. What is at the
heart of the physics are not questions of which boundary
conditions to take and how they should be modified, but rather how
measurement of physical quantities are made in general
relativity when the background metric varies with time.

When doing an experiment to measure some physical property of a
test particle\textemdash its phase shift, say\textemdash we start
by choosing an origin for our coordinate system. It is natural to
choose as this origin some point \textit{on} our measuring
apparatus, which in our case is the initial beam splitter. In the
three spatial dimensions, this origin is represented by a
\textit{point} on a three-dimension space; this point is what we use when
measurements with a \textit{physical} apparatus are actually made. If
we wish to represent the origin in space \textit{and} time, it would
seem to be natural to represent it as an \textit{event} in
four-dimensional \textit{spacetime}. Coordinate systems
constructed by doing so are called quasi-cartesian
coordinate systems by Synge. He also noted who also noted that a
suitable choice of origin in spacetime for is often difficult (see
section 10 of \cite{Synge}), however. This is especially true in the
case of gravitational waves, where the metric is nonstationary, and varies
with time. In fact, by identifying the origin as an
\textit{event}, we have ignored the fact that our measuring apparatus
is a \textit{physical} object. It too must travel along a
\textit{worldline} in the spacetime, as will any test particle whose
physical properties we are measuring. As was argued in \cite{GLF},
only \textit{relative} motions of a test particle with respect to the
apparatus can thus be measured. These arguments were based in part on the
following observation by Hawking and Ellis:
\begin{quote}
``In chapter 3 we saw that if the metric was static there was a
relation between the magnitude of the timelike Killing vector and
the Newtonian potential. One was able to tell whether a body was in a
gravitational field by whether, if released from rest, it would
accelerate with respect to the static frame defined by the Killing
vector. However, in general, space-time will not have any Killing
vectors. Thus one will not have any special frame against which to
measure acceleration; the best one can do is to take two bodies close
together and measure their \textit{relative acceleration} [emphasis
added].'' (From page 78 of \cite{Hawking}.)
\end{quote}
This relative motion is not measured in the TT coordinates where
the coordinates are defined relative to a fixed \textit{event},
chosen as the origin, in the spacetime. The motion of the
measuring apparatus through spacetime was not taking into account
in the TT coordinates, so that the beam splitters and the mirrors
are at \textit{fixed} coordinate values.  This relative motion, on
the other hand, is measured in the proper reference frame where
the coordinates are explicitly constructed to span the distance
between the measuring apparatus, and the test particle whose
properties are to be measured.  The motion of the measuring
apparatus has been built in from the beginning, and indeed, the
positions of the beam splitters and the mirrors shift during the
passage of the gravitational wave. Failing to take into account
the motion of the measuring apparatus in spacetime is the underlying
\textit{physical} reason why the wrong expression was obtain for
Fig.~$\ref{H-V-MIGO}$ when the TT coordinates are used.

These statements will be expanded upon, and justified in
\cite{ADS-Chiao-2004}.

\appendix

\section{The Quasi-classical Approximation and the Interferometer}

In this \textbf{appendix} we set the framework used in our different
calculations of the phase shift.

\subsection{The quasi-classical approximation}

Like Stodolsky \cite{Stodolsky, Alsing}, we use the
quasi-classical approximation to calculate the phase shift of an
atom in a matter-wave interferometer. This approximation is the
Lagrangian version of the WKB, or stationary phase approximation,
in the Schr\"odinger representation of quantum mechanics. We assume
that the density of particles through the interferometer is low
enough that they do not interact with one another appreciably in
the time it takes for the atom to traverse the interferometer.
Within this approximation, the particles  can be treated as test
particles that interact only with the gravitational wave, and not
with themselves. We can thus consider the phase shift of each atom
individually, and use the Feynman path integral method to
calculate its phase shift. However, instead of integrating over
all possible paths, we make the quasi-classical approximation, and
consider only the most \textit{probable} path\textemdash which is
also the \textit{classical} path\textemdash that the test particle
takes through the interferometer (see \cite{Berman}). As a result,
\begin{equation}
\Delta\phi \approx \frac{1}{\hbar}\bigg\{S_{cl}[x_A,x_B;\gamma_1]
  -S^{cl}[x_A,x_B;\gamma_2]\bigg\},
\label{def-phase-shift}
\end{equation}
where the superscript ``cl'' stands for ``classical'', and
$\gamma_1$ and $\gamma_2$ are the two possible paths a classical
test particle can take through the interferometer that links a
common point $x_A$ on the interferometer (the initial beam
splitter, say) to a common point $x_B$ on the interferometer (the
final beam splitter, say). Note that $S_{cl}[x_A,x_B;\gamma]$ are
functionals of the points $x_A$ and $x_B$, along with the paths
$\gamma_1$ and $\gamma_2$ connecting them.

Unlike \cite{Stodolsky}, as well as the calculations of the phase
shift done in \cite{Linet-Tourrenc, Papini, Borde1, Borde2,
Borde3, Alsing}, we will work in the nonrelativistic limit. The
reason for doing so is two fold. First, the proper reference
frame, as well as the general laboratory frame and all the other
reference frames based on the worldline of an observer, are
constructed for nonrelativistic test particles. Second, as is well
known, the phase shift of a matter-wave interferometer is
proportional to the mass of the particle. This argues for the use
of atoms in matter-wave interferometers to detect gravitational
waves, as was proposed in \cite{nmMIGO, CrysMIGO}, and in any
conceivable matter-wave interferometer, these atoms will be moving
nonrelativistically.

\subsection{The interferometer}

The configuration of the interferometer that we use in our
calculations is shown in Fig.~$\ref{H-V-MIGO}$. Details of this
interferometer can be found in \cite{nmMIGO}. For our purposes, it is
sufficient to note that the interfering test particle in the
interferometer is an atom with mass $m$. A beam of these atoms is
emitted from an atomic source with velocity $v_s$, and is
beam-split into different diffraction orders using a diffraction
grating with periodicity $a$ (see \cite{Pritchard, Borde4,
Zeilinger}). Only the $n=+1$ and $n=-1$ diffraction orders are
used in the interferometer.

There are two possible paths, denoted as $n=+1$ and $n=-1$ in
Fig.~$\ref{H-V-MIGO}$, that any individual atom can follow through the
interferometer; it is not possible to tell \textit{a priori} which of
the two paths an atom will take. If the atom is diffracted
into the $n=+1$ diffraction order, then immediately after the beam splitter
it will have a velocity of $+ v_\bot = 2\pi\hbar/ma$ perpendicular to
the longitudinal axis of the interferometer. Similarly, if an atom is
diffracted into the $n=-1$ diffraction order, it will have a velocity
of $- v_\bot$ (see Fig.~$\ref{H-V-MIGO}$). Since the diffraction
grating does not impart energy to the atom, for both diffraction
orders $v_\|=(v_s^2-v_\bot^2)^{1/2}$ is the velocity of the atom
parallel to the axis of the interferometer immediately after it leaves
the beam splitter.

In \cite{nmMIGO, CrysMIGO} the use of hard-wall, specular reflection
mirrors was considered for the matter-interferometer. Mirrors of this
type have been made from silicon wafers \cite{Holst}, and for near
normal and near glancing angles, the reflection from these mirrors is
close to 100\%. These are not the mirrors that are currently used in
matter-wave interferometry, however. Mirrors based on diffraction
gratings, standing waves of light, or stimulated Raman pulses are
used instead \cite{Pritchard, Borde4, Zeilinger, ChuKas,
Kas}. We will nevertheless consider only the case of hard-wall mirrors
in this paper, because of their simplicity. In the absence of gravitational
waves, the effect of hard-wall mirrors, if they do not move, on the
velocity of the atoms through the interferometer is
\begin{equation}
v_x(out)=v_x(in), \qquad v_y(out) = -v_y(in),
\label{bc}
\end{equation}
where $\vec{v}(in)$ and $\vec{v}(out)$ are, respectively, the
velocity of the atom immediately before it is reflected off the
mirror, and immediately afterwards.

\section{The TT gauge and the General Laboratory Frame}

In this appendix we will show how the TT gauge can be taken in the
general laboratory frame in a treatment that follows \cite{GLF}.

Like the proper reference frame, the general laboratory frame is
an orthonormal coordinate system fixed on the worldline of an
observer. In our case it is fixed on the worldline of the initial
beam splitter in Fig.~$\ref{H-V-MIGO}$. We choose a local tetrad
$e^A_{\>\>\>\>\mu}$ fixed on the observer's worldline. Thus,
$e^A_{\>\>\>\>\mu}e_{A\nu}= g_{\mu\nu}$, while
$e_A^{\>\>\>\>\mu}e_{B\nu}= \eta_{AB}$. In a slight break in the
notation followed elsewhere in this paper, which follows the
notational convention established in \cite{Thorne}, we use in this
\textbf{appendix} the convention taken in \cite{GLF}. Thus, we use
as the tetrad indices capital Latin letters, which run from 0 to 3.
Lower-case Latin letters are used as the spatial tetrad indices,
and they run from 1 to 3. As usual, Greek letters are used for the
spacetime indices of a general coordinate $x^\mu$.

Consider a the motion of a test particle $\cal P$ close to the
observer $\cal O$ and her worldline. Let $X^A(\tau)$ be the position
of $\mathcal P$ at any proper time $\tau$ in the observer's
frame. $X^A(x): x^\mu \to X^A$ can also be considered as a coordinate
transformation from the general coordinates $x^\mu$ to the tetrad
frame at any time $\tau$. Thus in a small neighborhood
$\mathcal{U}_\mathcal{O}$ of $\mathcal{O}$,
\begin{equation}
\eta_{AB}\> dX^A dX^B = \eta_{AB} \frac{\partial X^A}{\partial x^\mu}
\bigg\vert_{\mathcal{U}_{\mathcal{O}}} \frac{\partial X^B}{\partial
x^\nu}\bigg\vert_{\mathcal{U}_{\mathcal{O}}} dx^\mu dx^\nu,
\label{7}
\end{equation}
so that
\begin{equation}
\frac{\partial X^A}{\partial
x^\mu}\bigg\vert_{\mathcal{U}_{\mathcal{O}}} =
e^A_{\>\>\>\>\mu},\qquad \hbox{or}\qquad \frac{\partial
x^\mu}{\partial X^A}\bigg\vert_{\mathcal{U}_{\mathcal{O}}}
=e_A^{\>\>\>\>\mu}.
\label{8}
\end{equation}
Integrating Eq.~$(\ref{8})$, we find for the spatial coordinates
\begin{equation}
X_a(\tau) =
\int_{\mathit{\gamma}_\chi^\tau(0)}^{\mathit{\gamma}_\chi^\tau(s_1)}
e_{a\mu}(\tau,s)\chi^\mu(\tau,s) ds \equiv
\int_{\mathit{\gamma}_\chi^\tau(s)} \bm{e}_a.
\label{9}
\end{equation}
where $\bm{e}_A = e_{A\mu} dx^\mu$ is a 1-form, and $\gamma_\chi$ is a
space-like geodesic linking  $\cal O$ to $\cal P$ at a given proper
time $\tau$ of the observer. Next, after taking appropriate account of
causality, the time component is
\begin{equation}
X_0(t) = - \left(\tau-\tau'\right) +
\int_{\mathit{\gamma}_\pi(\sigma)} \bm{e}_0.
\label{11}
\end{equation}
where $\gamma_\pi$ is a null-vector linking $\cal O$ at time
$\tau'$ with $\cal P$ at time $\tau$; the difference $\tau-\tau'$
is the transit time of a light pulse emitted by $\cal O$, and
scattered by $\cal P$.

In the case of linearized gravity, $g_{\mu\nu} = \eta_{\mu\nu} +
h_{\mu\nu}$. It is straightforward to see that in terms of tetrads
\begin{equation}
e_A^{\>\>\>\>\mu} = \left(\delta^\mu_\nu - \frac{1}{2}h^\mu_\nu\right)
\delta_A^\nu, \quad \hbox{and }\qquad
e^A_{\>\>\>\>\mu} = \left(\delta_\mu^\nu + \frac{1}{2}h_\mu^\nu\right)
\delta^A_\nu.
\label{69}
\end{equation}
This choice is \textit{not} unique. If one does a \textit{local} Lorentz
transformation $L_A^{\>\>\>\>\tilde{A}}$ on
$e_A^{\>\>\>\>\mu}$, then $\eta_{AB} =
L_A^{\>\>\>\>\tilde{A}}L_{B\tilde{A}}$ while $e_A^{\>\>\>\>\mu} =
L_A^{\>\>\>\>\tilde{A}}e_{\tilde{A}}^{\>\>\>\>\mu}$. This leaves
$g_{\mu\nu}= e_{A\mu}\>\>e^A_{\>\>\>\>\nu}=e_{\tilde{A}\mu}\>\>
e^{\tilde{A}}_{\>\>\>\>\nu}$ invariant.

The usual derivation of the TT gauge makes use of a remnant of general
coordinate transformation invariance still left in the
linearized-gravity limit of general relativity. If $x_\mu =
\tilde{x}_\mu+\xi_\mu$ where $\xi_\mu$ is a `small', arbitrary vector,
then $h_{\mu\nu}$ transforms to
\begin{equation}
\tilde{h}_{\mu\nu} = h_{\mu\nu}+\partial_\mu\xi_\nu +
\partial_\nu\xi_\mu.
\label{trans}
\end{equation}
One can choose $\xi_\mu$ such that
\begin{equation}
\eta^{\mu\nu}h_{\mu\nu}=0,\qquad \partial^\mu h_{\mu\nu} = 0, \qquad
h_{0\mu} =0,
\label{TT-gauge}
\end{equation}
can be chosen for GWs in Minkowski space. Doing so defines a specific
coordinate system, which is called the TT coordinates in \cite{Thorne}.

The TT gauge can also be implemented in the general laboratory frame
as well. In terms of a tetrad frame, the linearized coordinate
transformation Eq.~$(\ref{trans})$ induces a Lorentz
transformation on the tetrads such that $e^{\tilde A}_{\>\>\>\>\mu} =
L^{\tilde A}_{\>\>\>\>A} e^B_{\>\>\>\>\mu}$, where
\begin{equation}
L^{\tilde A}_{\>\>\>\>A} = \delta^{\tilde A}_{\>\>\>\>A} +
\frac{1}{2}\left(\partial^{\tilde A}\xi_A +\partial_A\xi^{\tilde
A}\right), \label{Lor-Trans}
\end{equation}
and $\partial_A = e_A^\mu\partial_\mu$. Like the TT coordinates, the
TT gauge can also be realized, but now through the appropriate Lorentz
transformation. From Eq.~$(\ref{8})$, this in turn means we can choose
the TT gauge in the construction of the general laboratory frame as well.

\section{The High-frequency Limit}

In this \textbf{appendix} we consider the high-frequency
limit of $\Delta\phi_{GD}$.

Because of its dependence on three timescales, the analysis of the
high-frequency limit of $\Delta\phi_{GD}$ is more
involved than that for the low frequency limit. There are two cases to
consider. The simplest case is when the transit time of the atom
through the interferometer is very \textit{long} in comparison with
the period  of the gravitational wave, so that $T>> 2\pi/\omega$, but
$\omega<<\omega_0$ still. To the atom, the mirrors still do not move in
response to the gravitational wave, but the atom will experience many
oscillations of the gravitational wave as it traverses the
interferometer. In this limit the phase of the atom oscillates so fast
that we would naively expect the phase shift to average to zero. It
does not do so, however. We instead find that $F_{GD} \sim 1$.

That $\Delta\phi_{GD}$ does not vanish in the limit of
high-frequency gravitational waves is due to action of the mirrors
on the atoms as they traverse through the interferometer. When the
atoms reflect of the mirrors, they impart a impulsive force on the
atoms that is effectively \textit{instantaneous} \footnote{The
mirrors act on the atom over a timescale roughly proportional to
the size of the atom divided by $v_\bot$. Since this is much
shorter than any of the timescale considered here, the force can
be considered as instantaneous.}. The effect of this force is
readily apparent from the discontinuity in $v_y$ across the mirror
in Eqs.~$(\ref{bc})$ and $(\ref{pertGeobc})$. Note in particular
that the force is proportional to perturbed velocity
$v_{1y}(t_r+T^{-1}/2)$ of the atom \textit{caused by the
gravitational wave}, and is \textit{different} depending on
which path the atom travels through the interferometer, as can be seen
from the force lines in Fig.~$\ref{H-V-MIGO}$. It is because of this
net instantaneous force that $\Delta\phi_{GD}$ does not average to
zero in the high-frequency limit.

When, $\omega>>\omega_0$ as well, the mirrors themselves oscillate
rapidly as the atom traverses the interferometer. This rapid motion of
the mirror is not correlated with the motion of the atoms through the
interferometer, however. When the atom reaches the mirror, roughly
half the time it will be moving in the same direction as
the mirror, which would, from Eq.~$(\ref{GeoDevBC})$, tend to
\textit{decelerate} the atom. The other half of the time it
will be moving in the opposite direction as the mirror, and would tend
to \textit{accelerate} the atom. Not surprising, in this limit the
resonance function decreases by a half, and
$F_{GD} \approx 1/2$.

In this high-frequency limit $\Delta\phi_{GD}$ is
\textit{not} proportional to $R_{0x0y}$. There is no reason to expect
it to be, as there was in the low-frequency limit, however. Because
the mirrors exert a net force\textemdash which itself is the result of
the passage of the gravitational wave through the system\textemdash on
the atom as it passes through the interferometer, $\Delta\phi_{GD}\sim
\Gamma^x_{0y}$.

%\appendix
%\section{}

\begin{acknowledgments}

ADS and RYC were supported by a grant from the Office of Naval
Research.  We thank John Garrison, Jon Magne Leinaas, William Unruh,
and Rainer Weiss for clarifying and insightful discussions.

\end{acknowledgments}


\end{document}